\documentclass[showpacs,prd,amsmath,nofootinbib,amssymb]{revtex4}
\usepackage{graphicx,color}
\usepackage{subfig}
\usepackage{amssymb}
\usepackage{hyperref}
\usepackage[utf8]{inputenc}
\usepackage[english]{babel}
\usepackage{epsfig}
\usepackage{wasysym}
\usepackage{color,xcolor}
\usepackage{pdflscape}
\usepackage{amsmath}
\usepackage{bm}
\usepackage{epsfig}
\usepackage{amsfonts}
\usepackage{dcolumn}

\begin{document}

\title{Bardeen solution with a cloud of strings \\}

\author{Manuel E. Rodrigues$^{(1,2)}$\footnote{E-mail 
address: esialg@gmail.com}, Henrique A. Vieira$^{(1)}$\footnote{E-mail 
address: henriquefisica2017@gmail.com}
}

\affiliation{$^{(1)}$Faculdade de F\'{i}sica, Programa de P\'{o}s-Gradua\c{c}\~{a}o em F\'{i}sica, Universidade Federal do Par\'{a}, 66075-110, Bel\'{e}m, Par\'{a}, Brazil\\
$^{(2)}$Faculdade de Ci\^{e}ncias Exatas e Tecnologia, Universidade Federal do Par\'{a} Campus Universit\'{a}rio de Abaetetuba, 68440-000, Abaetetuba, Par\'{a}, 
Brazil\\
}

\begin{abstract}
In this paper we present a Bardeen solution surrounded by a cloud of strings fluid. We show how this model has the same event horizon characteristic as the Bardeen solution, however, the parameter of the strings make the solution singular at the origin. We also analyze the solution from a thermodynamic point of view. We calculate the system's state function, enthalpy; the temperature; and the other potentials; as a function of entropy. By analyzing the thermodynamic coefficients, we show that the solution presents three distinct phases, two of which are stable and one unstable. These three phases could be best visualized using a plot of Gibbs free energy versus temperature. In the end, we calculate the critical exponents and find that they are the same as those found in Van der Walls theory.

\end{abstract}

\date{\today}

\maketitle

\section{INTRODUCTION}
With the development of general relativity at the beginning of the last century \cite{Einstein1905}, we started to see the Universe in a completely different way. Schwarzschild \cite{SZ}, solving Einstein's equations, 
proposed the existence of a region of space time with singular density from which not even electromagnetic radiation could escape. Later, this region would become known as black holes, solutions to the Einstein equations that describe regions of space time covered by an event horizon and that may have a singularity inside \cite{Wlad}. This field challenged the understanding of physicists from that time and is still today subject to theoretical studies and motivation for the improvement of observational astronomical apparatus. 
And it has recently gained strength with the detection of gravitational waves \cite{LIGO1,LIGO2,LIGO3,LIGO4,LIGO5,LIGO6,LIGO7}, and with 
the release, by the Event Horizon Telescope team, of the first images of the shadow of a black hole \cite{fotoBN1,fotoBN2,fotoBN3,fotoBN4,fotoBN5,fotoBN6}.

Until the mid-1970s we thought it was impossible for any object to come out of a black hole, even electromagnetic radiation. However, Hawking \cite{SH2} showed that these bodies emit thermal radiation and this started a new area in gravitation known today as black hole thermodynamics \cite{bhtermo1,bhtermo2,bhtermo3,bhtermo4,bhtermo5, bhtermo6,bhtermo7,bhtermo8,bhtermo9,bhtermo10}. There are several reasons why this area is relevant today. First, any physicist knows the difficulty of trying to describe, for example, one gas using the equations of motion from mechanics. Now, imagine trying to evaluate the contribution of each star, or even each galaxy, separately, in the formation and functioning of the Universe. This is why thermodynamics is an essential part of the analysis of cosmology, which deals with distances of the order 1 mega parsec. Second, a purely classical black hole has no entropy and therefore would violate the 2nd law of thermodynamics. Last but not least, it is possible to construct purely classical laws for black holes \cite{Leis} that can only be interpreted by adding quantum effects, and the reason for this is not yet known since we do not have a quantum theory of gravity, neither one that can explain all forces of nature in a simple and condensed way.

Recently, the string theory is the most referred possibility to be “the final theory”. On it, the Universe is thought of as a collection of extended objects instead of pointlike particles. A promising candidate is a one-dimensional continua string object. 
 M. G\"{u}rses and F. G\"{u}rsey \cite{Gurses} first derived the string equation of motion in General Relativity, then they  showed \cite{Gurses2} that a fluid governed by this equation could model the interior of a Kerr-Schild metric. Later, J. Stachel \cite{Stachel} proposed an extension of the relativistic “dust cloud" model for a perfect fluid. With this in mind, Letelier \cite{Letelier} obtained a solution of Einstein's equations for clouds of strings and used it to construct a model of a star. Subsequently, many other papers in the literature have considered clouds of strings a fluid serving as a background for black holes, charged or not. In the Einstein-Gaus-Bonet theory: it was first considered on \cite{strings4}, where the authors also compute Hawking temperature, entropy, heat capacity, and Helmholtz free energy; then \cite{strings4b} calculated quasinormal modes for a scalar field in such space-time, highlighting the function of the parameter associated with the clouds of strings; at \cite{strings4c} they included charge, which could be a monopole, and focused on thermodynamic analysis. There are also solutions for higher orders considering Lovelock theory: Ghosh, Papnoi, and Maharaj \cite{strings2} found first to third-order solutions of $D \geq 4$ dimensions and they made thermodynamic stability considerations; following the previous work, Ghosh and Maharaj \cite{strings2b} generalized the solution to $N$ dimensions, evaluated the energy conditions, and studied the role played by clouds of strings in the structure of the event horizon. Another class of solutions involving clouds of strings are those that are also considered the so-called quintessence. Today, we have experimental data pointing to an accelerating expanding universe \cite{Perlmutter, Riess, Garnavich}, and yet no theoretical possibility has been widely accepted as the explanation for this phenomenon, with the ideas of matter and dark energy being the most widespread candidates within the literature. Quintessence is currently seen as the candidate to explain dark energy \cite{Kiselev} and thereby solve the question of the accelerating expansion of the Universe. Toledo and Bezerra \cite{strings6} find a Reissner–Nordström type solution then studied its thermodynamics and calculated the quasinormal frequencies for a scalar field; the same authors extended the previous work to higher dimensions using the Lovelock theory
\cite{strings2c}; on \cite{strings6b} we have an extension of Letellier space-time and is showed that this model corresponds to a black hole with global monopole surrounded by quintessence; finally, on \cite{strings6c} they presented a charged AdS black hole and studied it's thermodynamics aspects. Cai and Miao \cite{strings6d} have further shown, using Rastall gravitation, that it is possible to obtain a black hole solution where the clouds of strings become quintessence in a given limit.  It is also worth mentioning that Richarte and Simeone \cite{strings3} built a traversable wormhole model using Letelier's idea.

There are several types of black hole solutions today, however, the one proposed by James Bardeen\cite{Bardeen1} was revolutionary because there is no curvature singularity. Initially, the model lacked a source, but a few years later Beato and Garcia \cite{Beato}  proposed a self-gravitating magnetic charge, described by nonlinear electrodynamics \cite{Born},  as a source and this make it a exact solution of Einstein's equations.  It was realized that one can use nonlinear electrodynamics to construct regular black hole \cite{regular4}, and now we have solutions that:  possess electric charge as source \cite{regular2,regular5}; it's obtained using $f(r)$ theory \cite{regular1,regular3,regular9}, and also using $f(G)$ theory \cite{regular9,regular8}; possess angular momentum \cite{regular6,regular7}; and on Rainbow Gravity \cite{regular10}. There is also in the literature a Bardeen solution surrounded by the quintessence \cite{Henrique}, a work that shows how the regularity of the model is lost when a singular solution is attached to it. Therefore, we aim here to build a solution that is Bardeen surrounded by clouds of strings.

The paper is organized as follows: in section \ref{SECUM} we will comment on the general aspects of how we did to obtain our solution; in section \ref{SECDOIS} we will find the event horizons and make considerations about regularity; in section \ref{SECTRES} we will work on the solution from a thermodynamic point of view; and finally, in section \ref{SECQUATRO} we will make our final considerations.  We will consider throughout this work the metric signature $(+,-,-,-)$. The Riemann tensor is defined as $R^{\alpha}_{\ \beta \mu \nu} = \partial_{\mu} \Gamma^{\alpha}_{\ \beta \nu} -  \partial_{\nu} \Gamma^{\alpha}_{\ \beta \mu} + \Gamma^{\sigma}_{\ \beta \nu}\Gamma^{\alpha}_{\ \sigma \mu} - \Gamma^{\sigma}_{\ \beta \mu}\Gamma^{\alpha}_{\ \sigma \nu}$, where $\Gamma^{\alpha}_{\ \mu \nu}= \frac{1}{2} g^{\alpha \beta} \left(  \partial_{\mu} g_{\nu \beta}  + \partial_{\nu} g_{\mu \beta} 
- \partial_{\beta} g_{\mu \nu}  \right)$ is the Levi-Civita connection. Also, we will use geometrodynamic units where $G = \hbar = c =1$.

\section{SPACE TIME WITH CLOUDS OF STRINGS 
\label{SECUM}}
We are looking for a Bardeen solution in a spacetime surrounded by clouds of strings, this type of solution can be derived from general relativity minimally coupled to non linear electrodynamics (NED) and the cloud of strings by the action

\begin{equation}
    S = \int d^4x \sqrt{-g} \left[ R + 2 \lambda  + \mathcal{L}(F) \right] + S_{CS},
    \label{eq:acao}
\end{equation}
where $g$ is the metric determinant, $R$ is the curvature scalar, $\lambda$ is the cosmological constant, $\mathcal{L}(F)$  is the nonlinear general Lagrangian of electromagnetic theory, function of the scalar $F = F^{\mu \nu}F_{\mu \nu}/4$, and $S_{CS}$ is the Nambu–Goto action to used to describe stringlike objects, given by \cite{Letelier}

\begin{equation}
    S_{SC} = \int  \sqrt{- \gamma} \mathcal{M} d \Lambda^0 d \Lambda^1.
    \label{nambu}
\end{equation}

Where, $\gamma$ is the determinant of $\gamma_{AB}$, which is a induced
metric on a submanifold given by

\begin{equation}
    \gamma_{AB} = g_{\mu \nu} \frac{\partial x^{\mu}}{ \partial \Lambda^{A}}  \frac{\partial x^{\nu}}{ \partial \Lambda^{B}},
\end{equation}

 $\mathcal{M}$ is a dimensionless constant that characterize the string, and $\Lambda^0$ and $\Lambda^1$ are a timelike and spacelike parameter. Note that here each string is associated with a world sheet and is described by $x^{\mu} (\Lambda^{A})$. It's possible to rewrite \eqref{nambu} as
 
 \begin{equation}
      S_{SC} = \int  \mathcal{M} \left(   - \frac{1}{2} \Sigma^{\mu \nu} \Sigma_{\mu \nu} \right)  d \Lambda^0 d \Lambda^1,
 \end{equation}
 
 where $\Sigma^{\mu \nu}$ is a bi-vector written as
 \begin{equation}
     \Sigma^{\mu \nu} = \epsilon^{AB} \frac{\partial x^{\mu}}{ \partial \Lambda^{A}}  \frac{\partial x^{\nu}}{ \partial \Lambda^{B}},
 \end{equation}

note that $\epsilon^{AB}$ is the Levi--Civita symbol, $\epsilon^{01}= -\epsilon^{10}=1$. 

 Varying the action \eqref{eq:acao} with respect to the metric we find 
 
\begin{equation}
    R_{\mu \nu} - \frac{1}{2} g_{\mu \nu}R + g_{\mu \nu} \lambda = 8 \pi T_{\mu \nu} + 8 \pi T_{\mu \nu}^{SC},
    \label{eqeinsteinc}
\end{equation}

where $T_{\mu \nu}$ is the stress-energy tensor of the matter sector, defined for NED as
 \begin{equation}
T_{\mu \nu} = g_{\mu \nu} \mathcal{L}(F) - \frac{d \mathcal{L}}{d F} F_{\mu}^{\ \alpha} F_{\nu \alpha},
\end{equation}
 
 and $T_{\mu \nu}^{SC}$ is the stress-energy tensor of the cloud of string, which has the form \cite{Letelier}

 \begin{equation}
     T_{\mu \nu}^{SC} = \frac{\rho \Sigma_{\mu}^{\ \alpha} \Sigma_{\alpha \nu} }{8 \pi \sqrt{-\gamma}},
     \label{TEMN}
 \end{equation}
 with $\rho$ being the proper density of the cloud.
Requiring a zero divergence of the  stress-energy tensor \eqref{TEMN}, we get the equations

\begin{equation}
    \nabla_{\mu} ( \rho \Sigma^{\mu \nu} ) = \partial_{\mu} \left( \sqrt{-g}  \rho \Sigma^{\mu \nu} \right) =0,
\end{equation}
and
\begin{equation}
    \Sigma^{\mu \beta} \nabla_{\mu} \biggl[ \frac{ \Sigma_{\beta}^{\ \nu}}{(-\gamma)^{1/2}}   \biggr] = 0.
\end{equation}

For a spherically symmetric spacetime we have

\begin{equation}
   ds^2 = f(r) dt^2 - \frac{1}{f(r)} dr^2 -r^2 d \theta^2 - r^2 \sin^2{\theta} d \phi^2,
\end{equation}
we will have only one nonzero component of the $\Sigma^{\mu \nu}$ bivector which is $\Sigma^{01}$
and it depends exclusively on the radial coordinate. Solving the two previous equations for $\Sigma$, we get

\begin{equation}
    \Sigma^{01} = \sqrt{-\gamma} = \frac{a}{\rho r^2},
\end{equation}
where $a$ is an integration constant related to strings, being limited to the interval $0 < a < 1$ \cite{Letelier}.

Now, defining the Lagrangian for the Bardeen solution as
 \begin{equation}
    \mathcal{L}(F) = \frac{3}{8 \pi sq^2 } \left(   \frac{\sqrt{2 q^2 F}}{2+\sqrt{2 q^2 F}} \right)^{5/2},
    \label{lbardeen}
\end{equation}
where $s = |q|/(2 m)$, we can solve the Einstein equations  \eqref{eqeinsteinc} getting only two nontrivial differential equations
 
 \begin{equation}
     -\frac{a}{r^2}+\frac{-r f'(r)-f(r)+\lambda  r^2+1}{r^2}-\frac{6 M}{q^3 \left(\frac{r^2}{q^2}+1\right)^{5/2}} = 0,
 \end{equation}
 
 \begin{equation}
   -\frac{f''(r)}{2}-\frac{f'(r)}{r}+\lambda -\frac{3 M \sqrt{\frac{q^2}{r^2}+1} \left(-5 q^{10}+2 q^2 r^8+2
   r^{10}\right)}{q^3 r^2 \left(q^2+r^2\right)^4} = 0.
 \end{equation}
 
  Solving the system of differential equations above we have
 

\begin{equation}
f(r) =   1-a -\frac{2 M_1}{r} -\frac{2 M r^2}{\left(q^2+r^2\right)^{3/2}}-\frac{\lambda 
   r^2}{3},
   \label{eq:metricam1}
\end{equation}
where $M_1$ is a integration constant. In order to get a Bardeen-AdS solution for when $a=0$ we do from here $M_1 = 0$. So, we  have

\begin{equation}
f(r) =   1-a -\frac{2 M r^2}{\left(q^2+r^2\right)^{3/2}}-\frac{\lambda 
   r^2}{3}.
   \label{eq:metrica}
\end{equation}

 We see that if $a=0$ on \eqref{eq:metrica} the Bardeen's solution \cite{Beato} is recovered.

\section{EVENT HORIZONS AND REGULARITY \label{SECDOIS}}

\begin{figure}[htpb]
    \centering
    \includegraphics[scale=0.52]{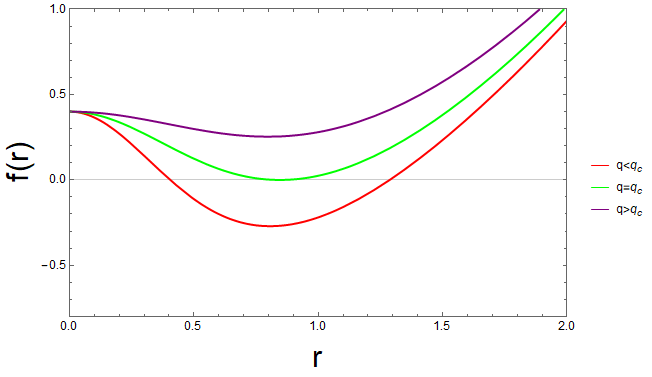}
    \caption{ Graphic representation of f(r) with respect to the radial coordinate with the values: $M=1$, $\lambda=-1$ and $a=0.6$. }
    \label{fig:BNum}
\end{figure}

To obtain the horizons, we need to solve $f(r) = 0$. Although we were not able to obtain an analytical expression, we show the behavior of $f(r)$ as a function of $r$ in the figure \ref{fig:BNum}. It is clear that the number of horizons depends on the values of the parameters, we see that it can be up to two. 

For any extreme solution we have a degeneracy in the event horizon, the condition to find the event horizon is, as said before

\begin{equation}
    f(r_+) = 0, 
\end{equation}
where $r_+$ represents the horizon radius. And furthermore, we must also impose
\begin{equation}
    \dfrac{d f(r_+)}{d r_+} = 0.
\end{equation}
From this
we can obtain both the value of the radius of the event horizon, and that of the critical charge $q_c$. For those values specified on the figure \ref{fig:BNum} we have $q_c=0.9985$.

To check for singularities we need to analyze the  Kretschmann scalar \cite{blackbounce6}, $K = R^{\mu \nu \alpha \beta}R_{\mu \nu \alpha \beta}$, which for this case is

\begin{equation}
K  = \frac{4}{3} \Biggl[\frac{3 a^2}{r^4}+\frac{2 a \left(\lambda
+\frac{6 M}{\left(q^2+r^2\right)^{3/2}}\right)}{r^2}  +\frac{6
   \lambda  M q^2 \left(4
   q^2-r^2\right)}{\left(q^2+r^2\right)^{7/2}}  
   +2 \lambda ^2+\frac{9 M^2 \left(-4 q^6 r^2+47 q^4 r^4-12 q^2
   r^6+8 q^8+4 r^8\right)}{\left(q^2+r^2\right)^7} \Biggr].
\label{eq:scalarK}
\end{equation}

We can expand this function in Taylor series to $r>>1$ and $r<<1$ around the value zero

\begin{equation}
    K (r \sim 0) = \frac{4 a^2}{r^4}+\frac{8 \left(3 m q \left(4 \lambda  q^2-3 a\right)+36 m^2+\lambda ^2 q^6\right)}{3
   q^6}+\frac{8 a \left(\lambda +\frac{6 m}{q^3}\right)}{3 r^2}-\frac{30 m r^2 \left(q \left(4 \lambda 
   q^2-a\right)+24 m\right)}{q^8} + \mathcal{O}(r^3),
   \label{eq:K0}
\end{equation}
\begin{equation}
    K (r \sim \infty) = \frac{8 \lambda ^2}{3}+\frac{8 a \lambda }{3 r^2}+ \mathcal{O}\left(\left(\frac{1}{r}\right)^3\right).
    \label{Kinfi}
\end{equation}
We see in \eqref{eq:K0} two terms proportional to $a^2/r^4$ and $a/r^2$ that account for the divergence of the scalar when $r \rightarrow 0$. It is clear that if $a=0$ we return to the Bardeen case where spacetime is regular everywhere. In general, the Kretschmann scalar  has the limits
\begin{equation}
    \begin{aligned}
    &\lim_{r \to 0} K = \infty, \\
    & \lim_{r \to \infty} K =  \frac{8 \lambda^2}{3}.
    \end{aligned}
    \label{eq:limitescalar}
\end{equation}

\section{THERMODYNAMICS \label{SECTRES}}
We begin our study of thermodynamics introducing the entropy of a black hole given by \cite{SH2}

\begin{equation}
    S = \frac{A}{4},
    \label{shalking}
\end{equation}
where  $A$ is the area of the horizon of the black hole  \cite{Wlad}. So, considering \eqref{eq:metrica} we have
\begin{equation}
    S = \pi r_+^2.
    \label{eq:s}
\end{equation}
We can obtain the mass from the condition $f(r_+) = 0$. From \eqref{eq:metrica} it follows that

\begin{equation}
  M =  \frac{\left(\pi  q^2+S\right)^{3/2} (-3 a+8 P S+3)}{6
   \sqrt{\pi } S}.
   \label{eq:massaems}
\end{equation}

We are considering that the function $M(S,q,a,P)$  play the role of enthalpy $H=H(S,P)$, this is the so-called extended phase space, where the cosmological constant is associated to a positive thermodynamic pressure  by \cite{pvcriticality}

\begin{equation}
    P = - \frac{\lambda}{8 \pi}.
\end{equation}

From the standard thermodynamics we know that the temperature $T$ and volume $V$ are obtained, respectively, by

\begin{equation}
    T = \frac{\partial M}{\partial S} = \frac{\sqrt{q^2+\frac{S}{\pi }} \left(S (-a+8 P S+1)+2 \pi  (a-1) q^2\right)}{4 S^2},
    \label{eq:tkems}
\end{equation}
\begin{equation}
    V = \frac{\partial M}{\partial P} =\frac{4 \left(\pi  q^2+S\right)^{3/2}}{3 \sqrt{\pi }}.
    \label{eq:volume}
\end{equation}

The other thermodynamic potentials, related to charge $A_q$ and the constant from clouds of strings $A_a$, are

\begin{equation}
    A_q = \frac{\partial M}{\partial q}=\frac{\sqrt{\pi } q \sqrt{\pi  q^2+S} (-3 a+8 P S+3)}{2 S},
\end{equation}
and
\begin{equation}
    A_a =\frac{\partial M}{\partial a} =-\frac{\left(\pi  q^2+S\right)^{3/2}}{2 \sqrt{\pi } S}. 
\end{equation}

We can find Smarr's formula through the properties of homogeneous functions \cite{Hankey}. Rewriting \eqref{eq:massaems} as

\begin{equation}
  M \left( l^bS,l^cq,l^dP,l^ha \right) =  \frac{\left(\pi l^{2c} q^2+l^bS\right)^{3/2} (-3 l^ha+8 l^dP l^bS+3)}{6
   \sqrt{\pi } l^bS}.
\end{equation}

In order to isolate $l$, the only values we can assume are: $c=b/2$, $d=-b$, $b=1$ and $h=0$. Therefore, we will have

\begin{equation}
     M \left( lS,l^{\frac{1}{2}}q,l^{-1}P,l^{0}a \right) =  l^{1/2} \left[ \frac{\left(\pi  q^2+S\right)^{3/2} (-3 a+8 P S+3)}{6
   \sqrt{\pi } S} \right],
\end{equation}

which means that the enthalpy of the black hole is a function with a degree of homogeneity $n=1/2$. Then the Smarr's formula is

\begin{equation}
    M = 2 \left( T S - PV \right) + A_q q,
\end{equation}

and so the first law is

\begin{equation}
    dM = T dS + A_q dq + A_a da + VdP.
\end{equation}

It is important to emphasize that  equation \eqref{eq:massaems} being the enthalpy of the black hole functions as the fundamental equation of the system. A simple check we can do is to test the so-called "cross derivatives". And, as we expected, one can easily verify that all of them will be true, i.e.

\begin{equation}
    \frac{\partial T}{\partial P} =   \frac{\partial V}{\partial S}; \ \ \  \frac{\partial T}{\partial q} =
     \frac{\partial A_q}{\partial S}; \ \ \  \frac{\partial T}{\partial a} =  \frac{\partial A_a}{\partial S}; \ \ \  \frac{\partial V}{\partial q} =  \frac{\partial A_q}{\partial S}; \ \ \  \frac{\partial A_q}{\partial a} =  \frac{\partial A_a}{\partial q}.
\end{equation}

\begin{figure}[htpb]
    \centering
    \includegraphics[scale=0.5]{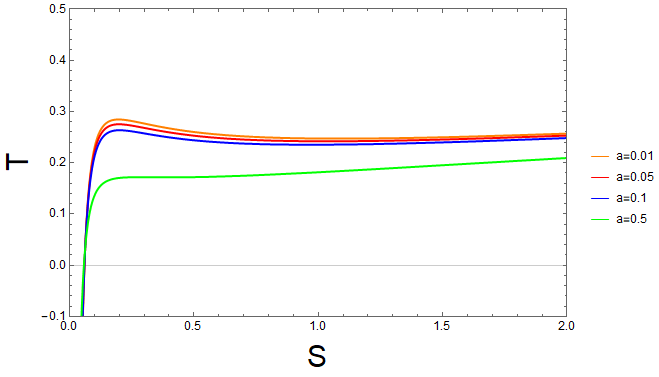}
    \caption{  Graphic representation of the temperature with the values: $\lambda=-0.08 \pi$ and $q=0.5$. }
    \label{fig:BNdois}
\end{figure}

\begin{figure}[htpb]
    \centering
    \includegraphics[scale=0.5]{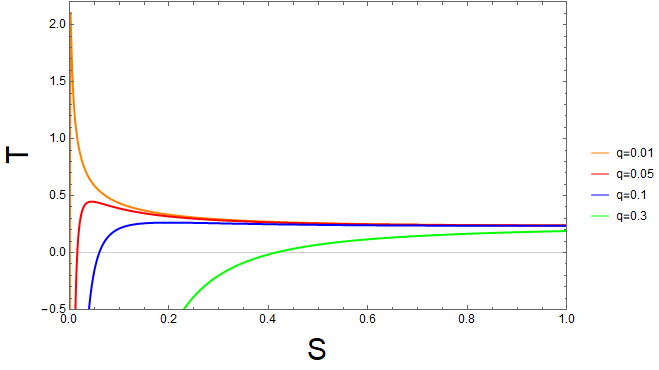}
    \caption{  Graphic representation of the temperature with the values: $\lambda=-0.08 \pi$, $a=0.6$ and $q=0.2$ (Bardeen curve). }
    \label{fig:BNtres}
\end{figure}

In figures \ref{fig:BNdois} and \ref{fig:BNtres} we show the behavior of temperature as a function of entropy. We realize that the parameter $a$ lower the temperature, and the same is true for charge. Also, we see that there is a minimum value of entropy, or radius of the event horizon, in order that temperature be positive. This negative region is tied to the factor $- 2 \pi q^2$ from \eqref{eq:tkems} and therefore we see increasing increments in the minimum entropy value as we raise the charge.

Going forward, we want to check the stability of the solution. To do that we need to analyze the $C_P$ and $K_T$ coefficients \cite{Stanley}, which are the heat capacity at constant pressure and the isothermal compressibility. They are given, respectively, by \cite{Davies}

\begin{equation}
    C_P = T \frac{ \partial S}{\partial T} \Bigg|_P,
\end{equation}
\begin{equation}
    K_T =  - \frac{1}{V} \frac{\partial V}{\partial P}  \Bigg|_T.
\end{equation}
 
From \eqref{eq:tkems} we have

\begin{equation}
    C_P = \frac{2 S \left(\pi  q^2+S\right) \left(S (-a+8 P S+1)+2 \pi
    (a-1) q^2\right)}{S^2 (a+8 P S-1)-4 \pi  (a-1) q^2 S-8
   \pi ^2 (a-1) q^4}
   \label{eq:CPBN}
\end{equation}

\begin{figure}[htpb]
    \centering
    \includegraphics[scale=0.5]{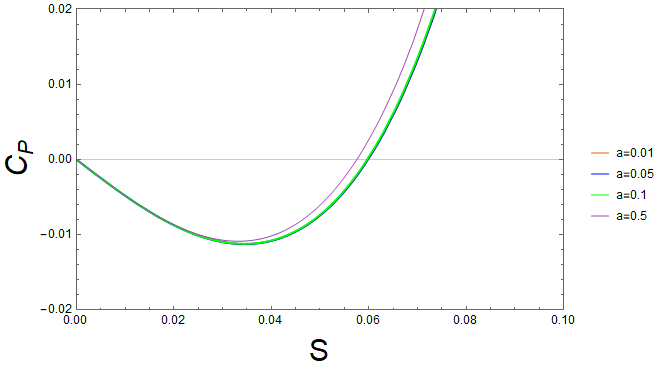}
    \caption{  Graphic representation of the heat capacity with the values: $P=0.1$ and $q=0.1$. }
    \label{fig:BNquatro}
\end{figure}

\begin{figure}[htpb]
    \centering
    \includegraphics[scale=0.5]{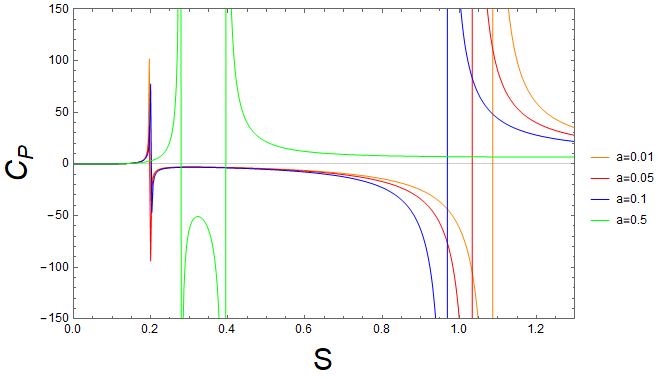}
    \caption{  Graphic representation of the heat capacity with the values: $P=0.1$ and $q=0.1$. }
    \label{fig:BNcinco}
\end{figure}

\begin{figure}[htpb]
    \centering
    \includegraphics[scale=0.5]{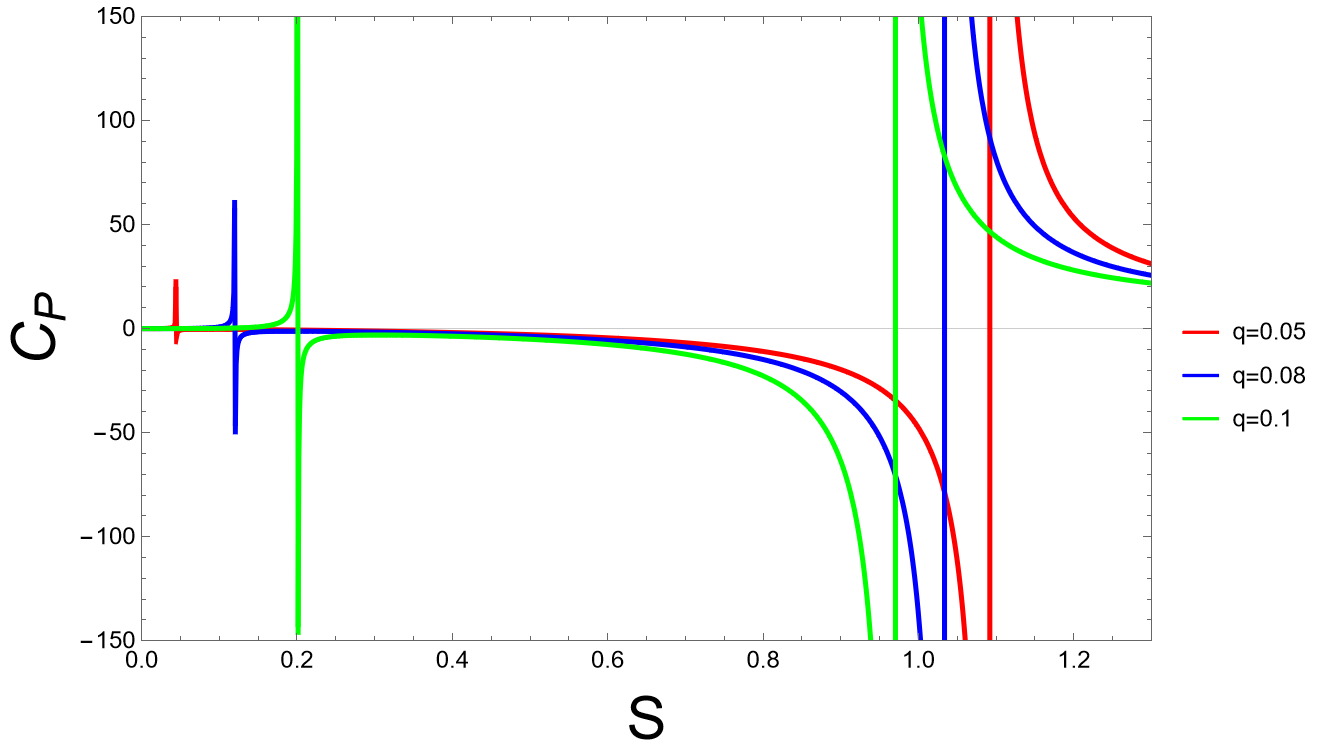}
    \caption{Graphic representation of the heat capacity with the values: $P=0.1$ and $a=0.1$.}
    \label{fig:BNseis}
\end{figure}

This expression can take on either positive or negative values, that is, there will be an alternation between stable and unstable regions. This is different from the case of Schwarzschild where $C_P = -2S$, and of curse this value is obtained when we do $q=0$, $\lambda = 0$ and $a=0$ on equation \eqref{eq:CPBN}. The divergences in heat capacity represent a phase transition, so we search for values of $S$ where the denominator of \eqref{eq:CPBN} is zero. In figures \ref{fig:BNquatro} and \ref{fig:BNcinco}  we show two regions with different entropy values, but with the same parameters $P=0.1$ and $q=0.1$. We see that there are two values of S where $C_P$ diverges, which means that there are two phase transitions. Since there is a correlation between entropy and event horizon radius, the phases of such a black hole are often referred on literature to as: small, small-large and large. 
In the first image, we see again that the solution has negative temperature values for $S<S_{min}$. In the figure \ref{fig:BNseis} we varied the charge and noticed that both parameters $a$ and $q$ tend to decrease the phase interval.

In order to find isothermal compressibility we need first obtain the state equation $P(T,V)$.  From 
\eqref{eq:tkems} and \eqref{eq:volume} we find the equation state, i.e

\begin{equation}
    P(T,V) = \frac{6 \sqrt[3]{6 \pi^2 V} \left(T V-(a-1) q^2\right)  +3 (a-1) V 
    +16 \pi ^2 q^4 T-8\ \sqrt[3]{36 \pi^4 V^2} q^2 T}{16 \sqrt[3]{6 \pi ^5 V}  q^4  
   -48 \pi  q^2 V+6\  \sqrt[3]{36 \pi V^5} }.
   \label{eq:estado}
\end{equation}

So, we will have

\begin{equation}
   \begin{aligned}
   K_T  = & \Biggl[ 3 \sqrt[3]{\pi V }  \left(-24  q^2 \sqrt[3]{\pi^2 V^2} +8
   \sqrt[3]{6 \pi^4}  q^4+3\sqrt[3]{36V^4} \right)^2 \Biggr] \times 
   \Biggl[  54 \sqrt[3]{\pi^2 V^5} \left(T V-4 (a-1) q^2\right)+24 \sqrt[3]{6 \pi^4}  q^2 V \bigl(5 (a-1) q^2  \\
  & -6 T V\bigr) +9 V^2 \sqrt{36V}  (a-1)
   -384 \pi^2 \sqrt[3]{(\pi V)^2} q^6 T  
  +144 \sqrt[3]{36V} \pi ^2 q^4 T V  +64 \sqrt[3]{6\pi} \pi^3 q^8 T  \Biggr]^{-1},
   \end{aligned}
   \label{eq:KTBN}
\end{equation}
\begin{figure}[htpb]
    \centering
    \includegraphics[scale=0.43]{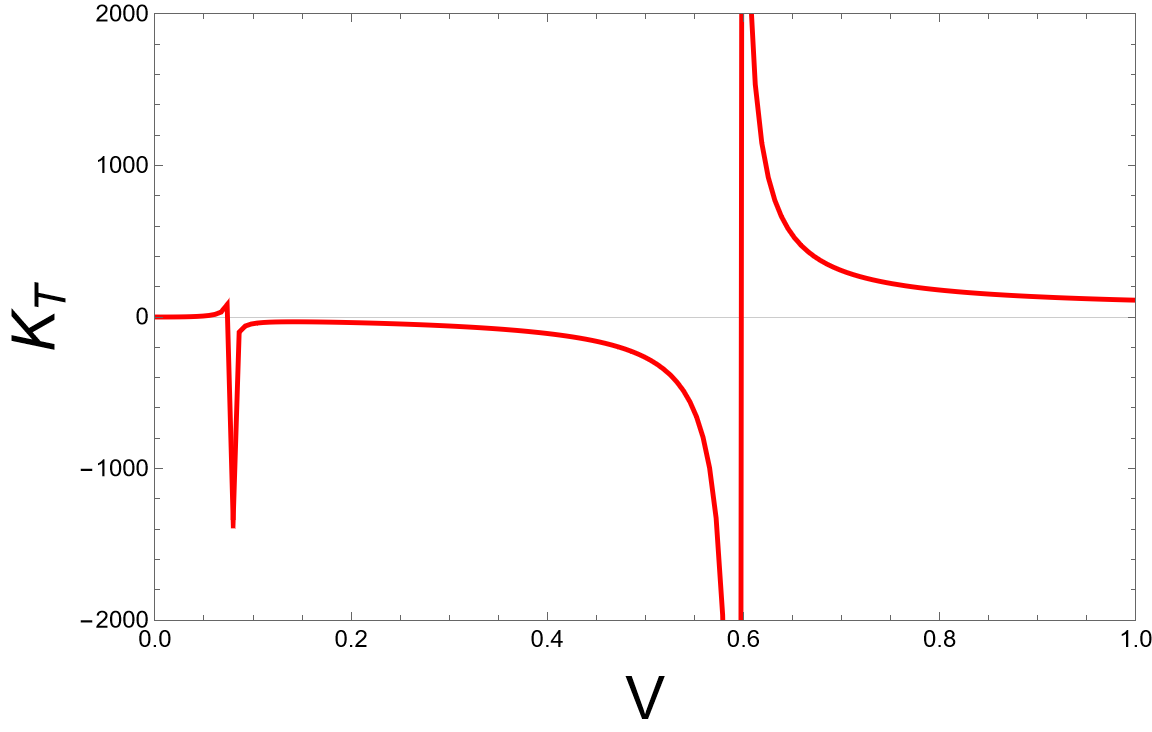}
    \caption{  Graphic representation of the isothermal compressibility with the values: $a=0.1$, $q=0.1$ and $T=0.8T_c$. }
    \label{fig:BNsete}
\end{figure}
which indicates the same aspects already mentioned in the heat capacity analysis, as we can see in the figure
\ref{fig:BNsete}. The temperature $T_c$ mentioned in the figure is the critical point obtained from the expression \eqref{eq:estado} by doing

\begin{equation}
    \frac{\partial P}{ \partial V} = 0 \ \ \ \ \text{and} \ \ \ \ \frac{\partial^2 P}{\partial V^2}.
    \label{eq:pontocritico}
\end{equation}

The critical values of temperature $T_c$, volume $V_c$ and pressure $P_c$ are given, respectively, by 

\begin{equation}
    T_c= \frac{5 \sqrt{188 \sqrt{10}-505} (1-a)}{432 \pi  q},
\end{equation}
\begin{equation}
    V_c = \frac{4}{81} \left(55+17 \sqrt{10}\right) \sqrt{188 \sqrt{10}-505} \pi 
   q^3,
\end{equation}
\begin{equation}
    P_c = \frac{0.00207151 (1-a)}{q^2}.
\end{equation}

\begin{figure}[htpb]
    \centering
    \includegraphics[scale=0.5]{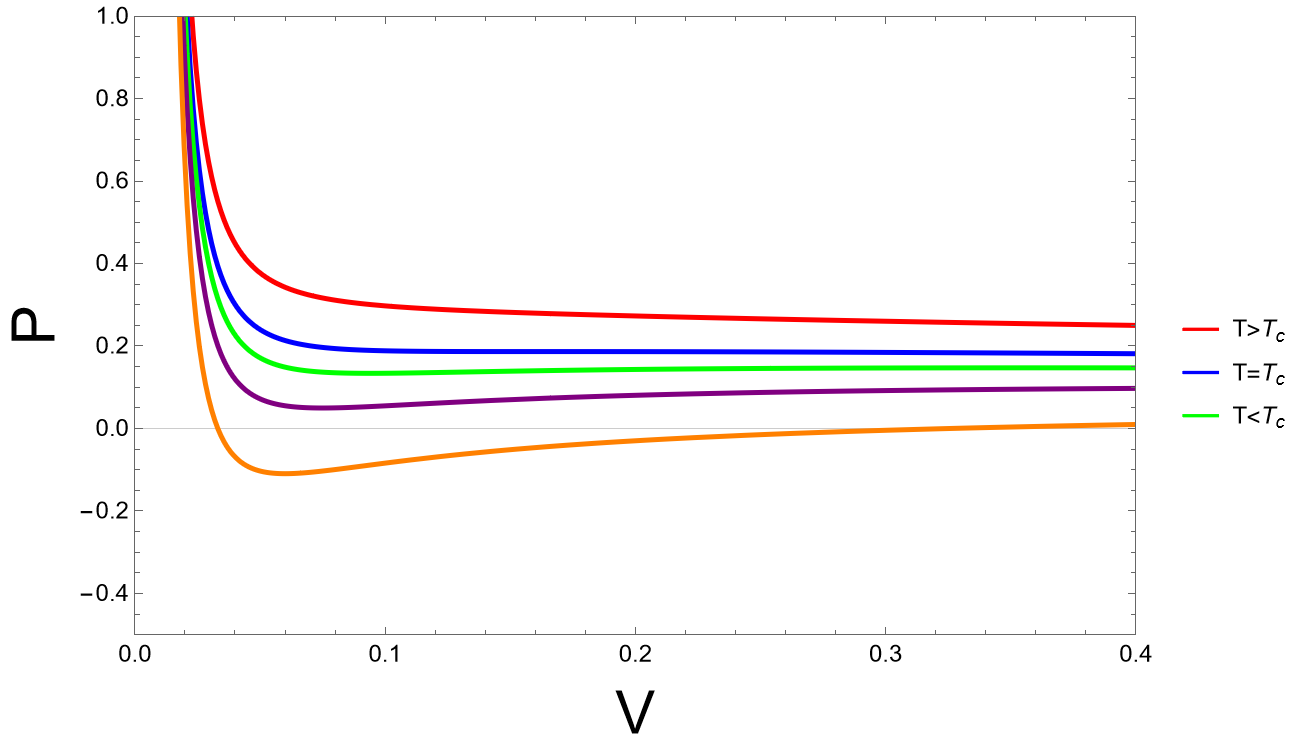}
    \caption{  Graphic representation of the state equation with the values: $a=0.1$, $q=0.1$. }
    \label{fig:oito}
\end{figure}

\begin{figure}[htpb]
    \centering
    \includegraphics[scale=0.5]{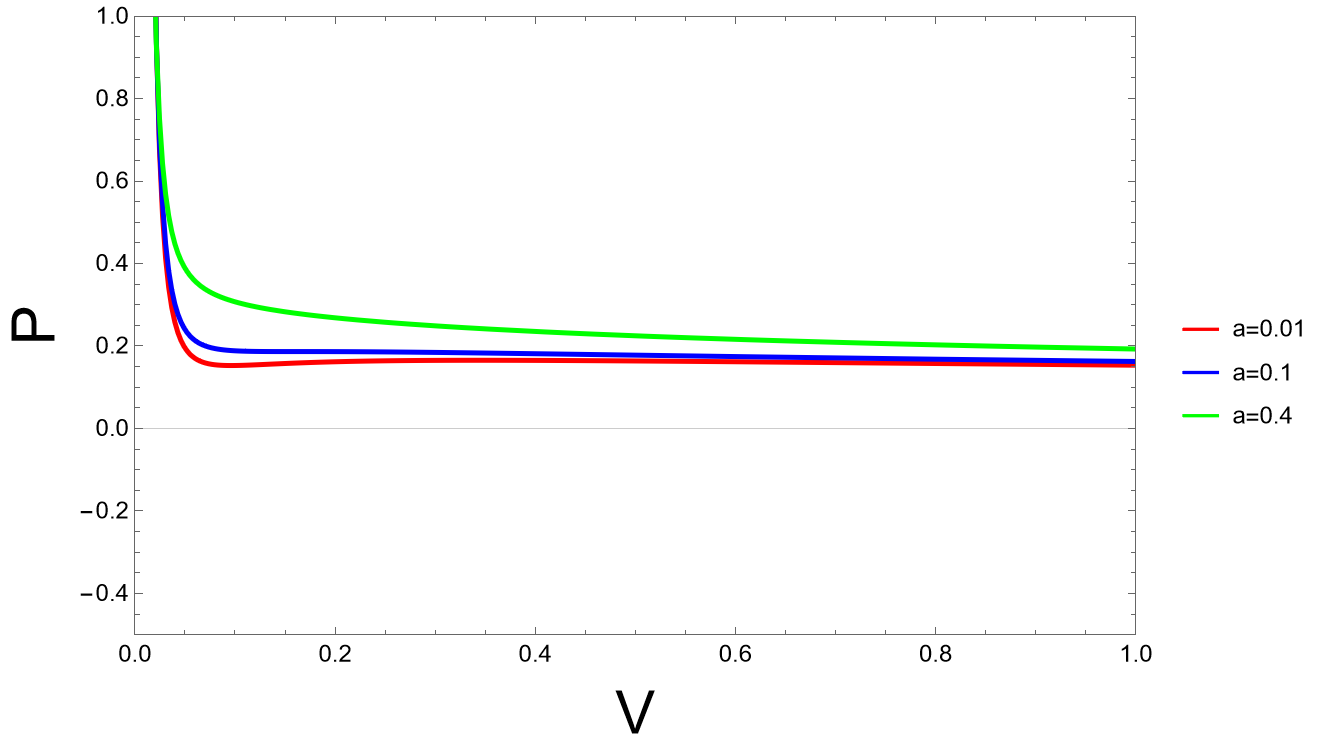}
    \caption{  Graphic representation of the state equation with the values: $T=T_c$ and $q=0.1$. }
    \label{fig:nove}
\end{figure}

\begin{figure}[htpb]
    \centering
    \includegraphics[scale=0.5]{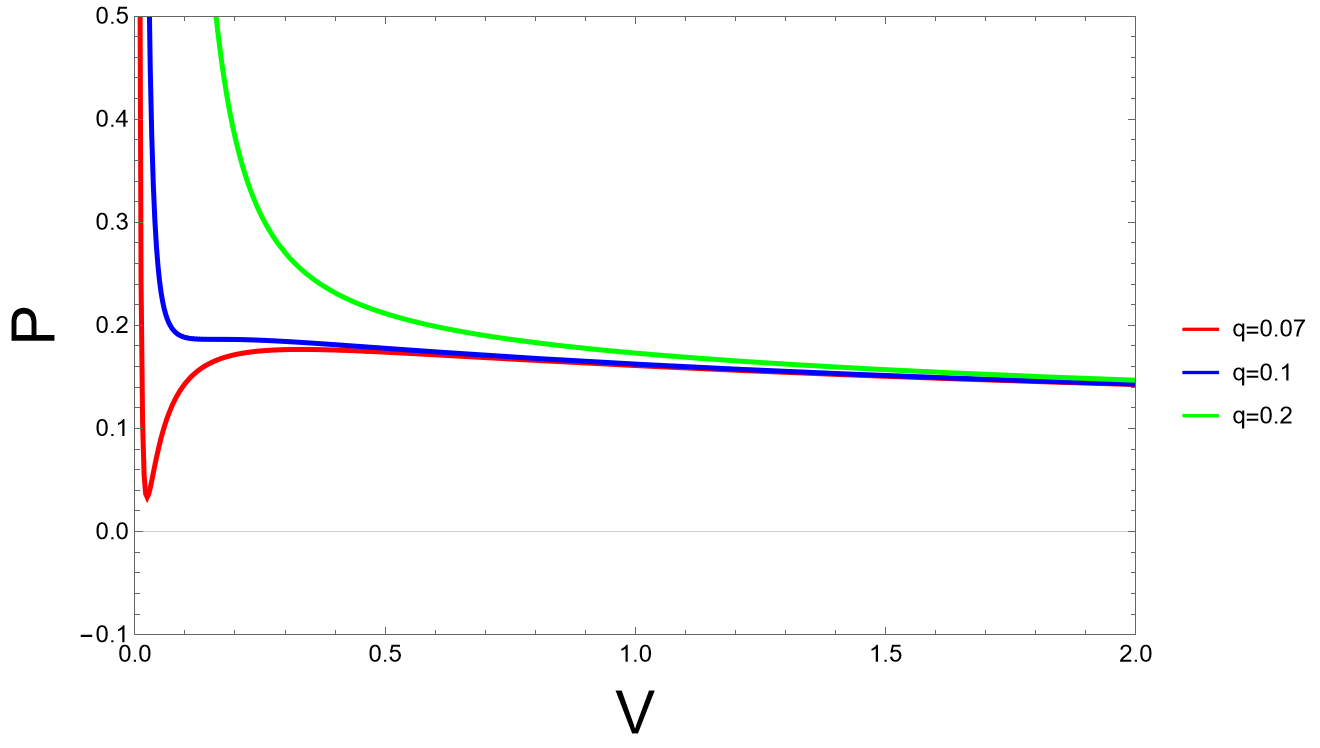}
    \caption{  Graphic representation of the state equation with the values: $a=0.1$ and $T=T_c$. }
    \label{fig:dez}
\end{figure}

Below, we show three images representing pressure as a function of volume. We see in figure \ref{fig:oito} that the solution has isotherms similar to the Van der Walls theory \cite{johnston}. Additionally, in \ref{fig:nove} and \ref{fig:dez}, we see that the pressure is lower for higher values of $a$ and $q$ parameters.

Having the equation of state, we can define the compressibility factor

\begin{equation}
    Z(T,V) = \frac{PV}{T}.
\end{equation}
This quantity indicates how much a gas deviates from ideal gases, which have $Z=1$ since there is no interaction between particles in it \cite{Stanley}. The compressibility factor for our case is

\begin{equation}
    Z(T,V) =  \frac{V \Bigl( 6 \sqrt[3]{6\pi^2V} \left(T V-(a-1) q^2\right)+3 (a-1) V 
    +16 \pi ^2 q^4 T -8 \pi \sqrt[3]{36 \pi V^2}q^2 T \Bigr)}{T \Bigl(16 \pi^2 \sqrt[3]{6 \pi V}q^4 -48 \pi  q^2 V+6 \sqrt[3]{36 \pi V } V^2\Bigr)},
   \label{eq:fatorz}
\end{equation}
and at the critical point
\begin{equation}
    Z_c = \frac{P_c V_c}{T_c} = 9.48724 q^2.
    \label{eq:zc}
\end{equation}
The function on \eqref{eq:fatorz} does not reproduce the same behavior as real gases because
$ Z \rightarrow \infty$ when $V \rightarrow \infty$ and of course it is expected that the limit value of $Z$ is 1. Furthermore, the compressibility factor given by \eqref{eq:zc} depends on the charge, in contrast to the Van der Walls case where it is independent of the interaction parameters.

Through a Legendre transformation on the enthalpy we obtain the Gibbs free energy \cite{Stanley}
\begin{equation}
    G = H - T S,
    \label{Fcorrigida}
\end{equation}
 from this function we can analyze the regions where each phase of the black hole presents itself. From \eqref{eq:tkems} and \eqref{eq:massaems} we have

\begin{equation}
G = \frac{\sqrt{q^2+r^2} \left(4 q^2 \left(-3a +4 \pi  P r^2+3\right)+r^2 \left(-3a -8 \pi  P r^2+3\right)\right)}{12
   r^2}.
\end{equation}

\begin{figure}[htpb]
    \centering
    \includegraphics[scale=0.5]{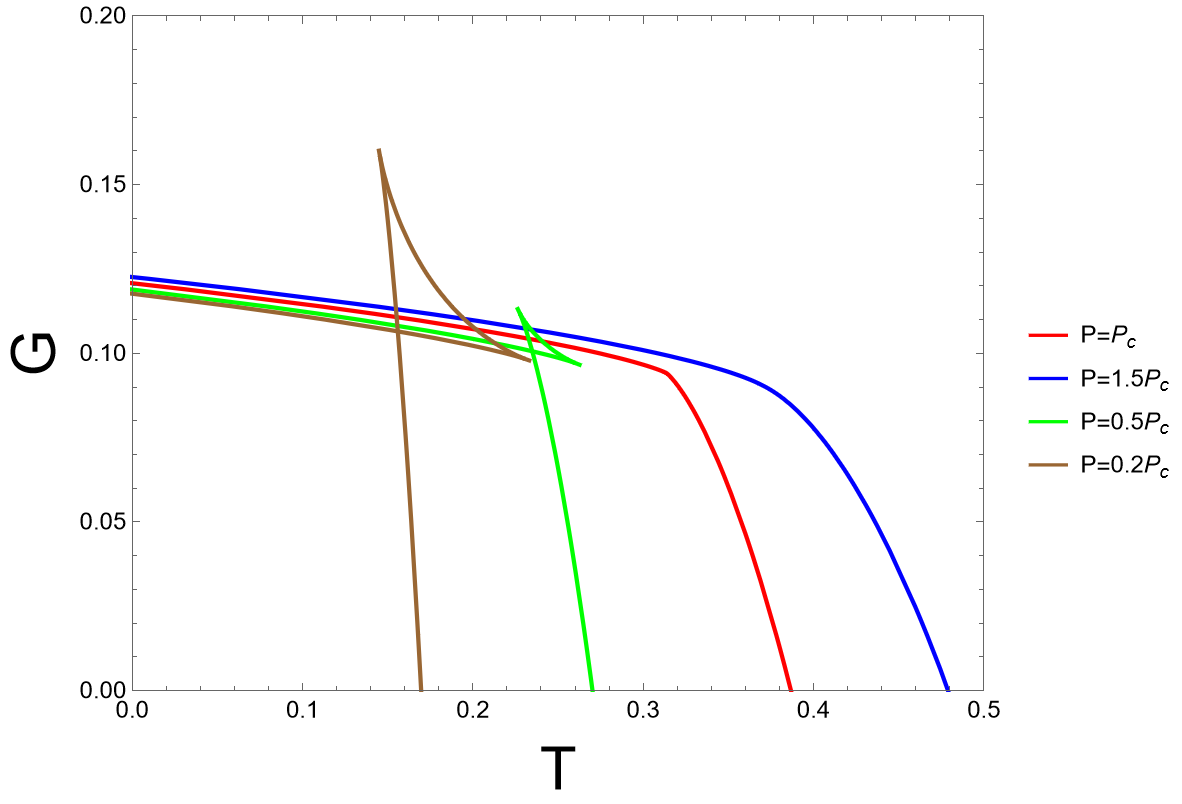}
    \caption{  Gibbs free energy as a function of temperature  with the values: $q=0.1$ and $a=0.1$. }
    \label{fig:15}
\end{figure}

\begin{figure}[htpb]
    \centering
    \includegraphics[scale=0.5]{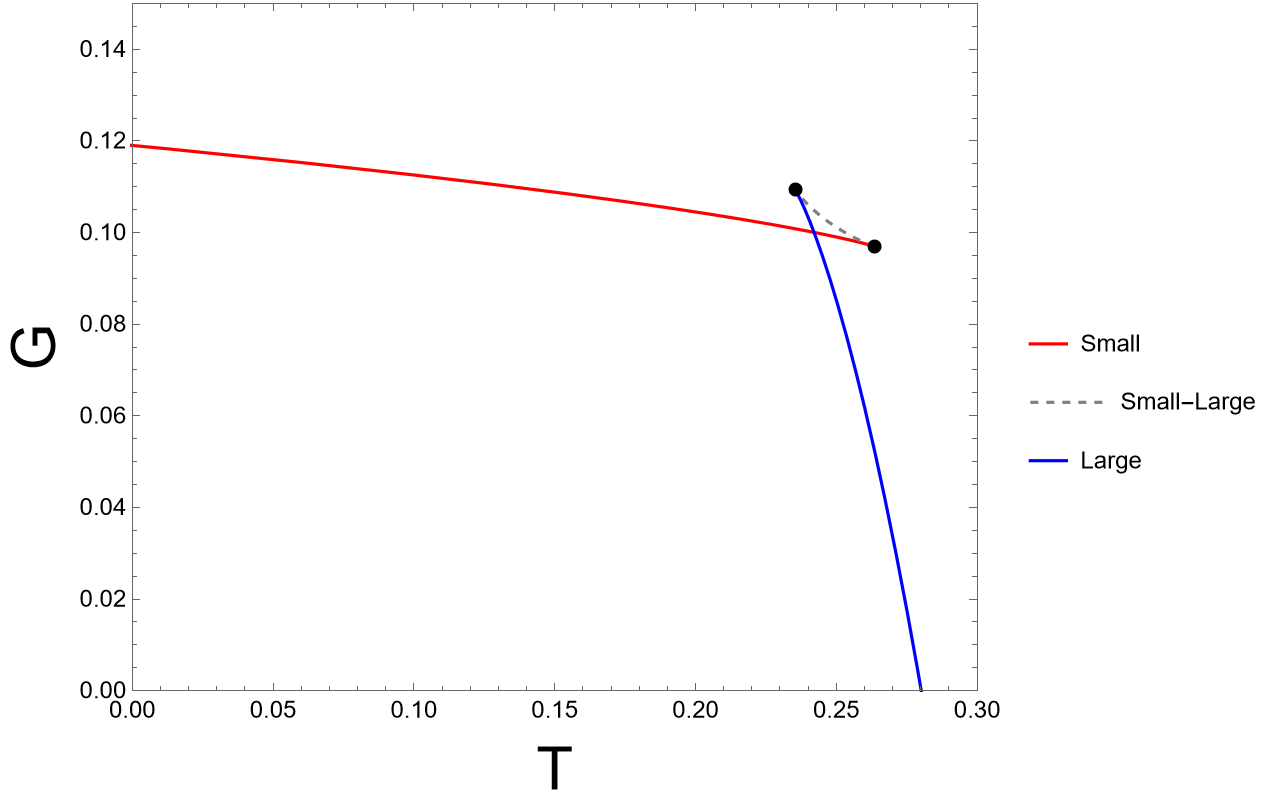}
    \caption{  Gibbs free energy as a function of temperature showing the phases of the Black Hole.  with the values: $P=0.1$, $q=0.1$ and $a=0.1$. }
    \label{fig:18}
\end{figure}

We notice in the figure \ref{fig:15} the characteristic behavior of a first-order phase transition  for values of $P<P_c$ and also points where a temperature value has up to 3 values of energy correspondents. Each value of free energy corresponds to a different phase of the black hole, as we can see in the figure \ref{fig:18}.

To finish our analysis we will calculate the critical exponents that describe the behavior of various thermodynamic functions near the critical point \cite{Stanley}. To obtain then, we use the so-called reduced variables
\begin{equation}
t=\frac{T-T_c}{T_c}, \ v=\frac{V-V_c}{V_c}, \ \mbox{and} \ p=\frac{P}{P_c}.\label{pvt}
\end{equation}

The  critical exponents are $\alpha$, $\beta$, $\gamma$, and $\delta$, defined, respectively, by
\begin{eqnarray}
C_V=T\left(\frac{\partial S}{\partial T}\right) \propto \left|t\right|^{-\alpha},\\
\eta=V_1-V_2 \propto \left|t\right|^{\beta},\\
k_T=-\frac{1}{V}\frac{\partial V}{\partial P}\propto \left|t\right|^{-\gamma},\\
\left|P-P_c\right|\propto \left|V-V_c\right|^{\delta},
\end{eqnarray}
where $\eta$ is the difference in volume between two phases. In our next steps we are assuming the values of the constants to be $q=0.1$ and $a=0.1$. If we use \eqref{pvt} and rewrite \eqref{eq:estado} and expand it for small values of $t$ and $v$, we get
\begin{equation}
p=1+t (B_1-B_2 v)-B_3 v^3+O\left(v^4,tv^2\right),\label{paprox}
\end{equation}
where $B_1$, $B_2$, and $B_3$ are just constants that are combinations of $q$ and $a$. If we consider $t$ as constant and derive $p$ to $v$, we get
\begin{equation}
dp=-\left(B_2t+3B_3v^2\right)dv.
\end{equation}
This relation is important to apply Maxwell's area law, which states \cite{Dayyani,WeiLiu}
\begin{equation}
\int VdP=0.
\end{equation}
Thus, considering Maxwell's area law and the fact that the pressure is constant at the phase transition, we obtain
\begin{eqnarray}
p&=&1+t (B_1-B_2 v_1)-B_3 v_1^3=1+t (B_1-B_2 v_2)-B_3 v_2^3,\\
0&=&\int_{v_1}^{v_2}\left(v+1\right)\left(B_2t+3B_3v^2\right)dv.
\end{eqnarray}
Solving these equations yields the nontrivial solution as follows:
\begin{equation}
v_1=-v_2 \propto \sqrt{t} \ \mbox{and} \ \eta\propto\sqrt{t}.
\end{equation}
This means that one of the critical exponents is $\beta=1/2$.

We can also use \eqref{paprox} to calculate the isothermal compressibility and we find that
\begin{equation}
k_T\propto \frac{1}{t},
\end{equation}
such that $\gamma=1$. 

We can rewrite the condition $\left|P-P_c\right|\propto \left|V-V_c\right|^{\delta}$ as $\left|p-1\right|\propto \left|v\right|^{\delta}$. From \eqref{paprox} we get
\begin{equation*}
p-1\propto t (B_1-B_2 v)-B_3 v^3.
\end{equation*}
However, in the critical isotherm we have $t=0$, so that,
\begin{equation}
p-1\propto v^3.
\end{equation}
With this result, we find the critical exponent $\delta=3$. Now, if we invert the equation  \eqref{eq:volume}, we find

\begin{equation}
    S(V,q) =  \frac{1}{4} \left( \sqrt[3]{36 \pi V^2 } -4 \pi  q^2\right),
\end{equation}

therefore, since entropy does not explicitly depend on temperature, we will have $C_V = 0$ which implies $\alpha=0$.

We see that these exponents satisfy the Griffiths, Rushbrooke and Widom equalities \cite{Stanley,WeiLiu,Griffiths};
\begin{eqnarray}
\alpha+\beta\left(1+\delta\right)=2,\ \mbox{Griffiths}\\
\gamma\left(\delta+1\right)=\left(2-\alpha\right)\left(\delta-1\right),\ \mbox{Griffiths}\\
\alpha+2\beta+\gamma=2, \ \mbox{Rushbrooke}\\
\gamma=\beta \left(\delta-1\right),\ \mbox{Widom}
\end{eqnarray}
which says that there are only two independent exponents.

\section{CONCLUSION \label{SECQUATRO}}
In this paper, we proposed a solution to Einstein's equations consisting of a Bardeen black hole surrounded by clouds of strings. We start from the modified action for nonlinear electrodynamics, as is done for Bardeen's solution, and add the Nambu-Goto action, which dictates the dynamics of the clouds of strings. The resulting metric is singular
due to the parameter $a$, as we found through analysis of the  Kretschmann scalar. Moreover, because of the result in \eqref{eq:limitescalar}, at infinity spacetime is asymptotically non-flat. We also find those event horizons have the same characteristic as Bardeen's solution, i.e., there are up to two for 
a charge smaller than the critical $q \leq q_c$.

We follow the idea of \cite{pvcriticality} and consider the black hole mass equation to work as a fundamental equation. Given the variables 
involved (entropy and pressure), we conclude that this is the enthalpy of the black hole. From this ansatz, we were able to perform a complete analysis from a thermodynamic point of view.  We find the first derivatives of enthalpy, that is, the so-called thermodynamic potentials, two of which, temperature 
and volume, functions that are highly relevant from a thermodynamic point of view.  We show that in fact, \eqref{eq:massaems} is a fundamental equation because from the 
potentials we calculate Smarr's formula and the first law of thermodynamics. We further strengthened our conviction by checking through the cross derivatives
that once again guaranteed the exact differential character of the mass. Through the analysis of the images \ref{fig:BNdois} and \ref{fig:BNtres} we detected that the temperature 
can assume negative values associated with the term $-2 \pi q^2$. However, we do not consider that regions with negative temperatures in this work.

As is usual in black hole thermodynamics, we also try to check the stability of our solution. Through the thermodynamic coefficients,
heat capacity at constant pressure \eqref{eq:CPBN} and isothermal compressibility \eqref{eq:KTBN}, we find that the solution is stable in two regions, separated by an unstable interval. The separation between these regions is given by a discontinuity, which indicates a first-order phase transition. Following the nomenclature adopted in the literature
we will call the black hole phases: small, small-large, and large. This is because the limit of each phase is related to entropy and consequently to the radius of the event horizon.  The phenomenological Van der Walls theory can partially explain phase transitions in classical thermodynamics, and in fact 
when we analyzed the equation of state of our solution \eqref{eq:estado} we see that it resembles Van der Walls isotherms.  We were even able to find
the critical point, characteristic of a second-order phase transition, which in our case represents the beginning of the coexistence region between the
small and large phases.  However, the compressibility factor did not reproduce the same characteristic of real gases, its value tending to infinity when the pressure 
approaches zero.

As a further confirmation of the existence of three distinct phases, we calculate the Gibbs free energy $G(T,P)$ and plot a graph of $G \times T$. We see in \ref{fig:15} that
a temperature point can have up to three corresponding energy values. This is also a characteristic that is present in Van der Walls gas and indicates
a region of phase correspondence.  As we know from standard thermodynamics, the free energy will be the minimum possible, the other values being unstable configurations of the system for that temperature. Finally, in order to get more information about the criticality of the solution we calculated the critical exponents. Again, the values concurred with the Van der Walls theory, they are $\beta = 1/2$, $\gamma=1$, $\delta =3$ and $\alpha= 0$.

We know that it is still possible to go further in this analysis, such as trying to use the methods of geometric thermodynamics to verify if the results
will be the same. However, we will leave this and other topics, such as the extrapolation to higher orders using Einstein-Gauss-Bonnet  or Lovelock theory and geodesic analysis, for future work.

 
\vspace{1cm}

{\bf Acknowledgements}: M. E. R.  thanks Conselho Nacional de Desenvolvimento Cient\'ifico e Tecnol\'ogico - CNPq, Brazil, for partial financial support. This study was financed in part by the Coordena\c{c}\~{a}o de Aperfei\c{c}oamento de Pessoal de N\'{i}vel Superior - Brasil (CAPES) - Finance Code 001.


\end{document}